# Improving Optical Metrology by Engineering the Target Environment

**Thomas A. Grant**[1†], **Cheng-Hung Chi**[1†], **Kevin F. MacDonald**[1*], **and Nikolay I. Zheludev**[1,2]

*1. Optoelectronics Research Centre, University of Southampton, Highfield, Southampton, SO17 1BJ, UK*

*2. Hagler Institute for Advanced Study, Texas A&M University, College Station, Texas, 77843, USA*

† Contributed equally to this work

* Correspondence to: kfm@orc.soton.ac.uk

**Abstract:** Measurements of positional coordinates and dimensions – whether by human vision or optical instrumentation – are fundamental to safety, industrial productivity, manufacturing quality/accuracy, and scientific discovery. The ultimate precision of such measurements is governed by the Fisher information conveyed from an object to a detector through the optical field, and strategies for enhancing measurement performance often focus on reducing detector noise and/or refining estimation algorithms. Building on the emerging understanding of Fisher information as a physical quantity that propagates through space in a wave-like fashion, we demonstrate that substantial gains in precision can also be made by engineering the electromagnetic environment of a measurement target to optimise the generation and transmission of Fisher information. Using nanowire position metrology based on light scattering at a wavelength $\lambda$ = 640 nm as an architype system, we achieve a multifold enhancement in localisation precision, reaching beyond $\lambda$/10,000. Our results establish target environment engineering as a powerful and broadly applicable strategy for advancing measurement and sensing performance across platforms ranging from optical characterisation of micro- and nano-objects to microwave radars and optical LiDAR navigation systems.

## Introduction

In optical metrology the ultimate limit on measurement precision is not set by the diffraction limit, but by measurement noise[1-4]. This limit is formalised by the Cramér-Rao lower bound, which states that the minimum achievable variance is determined by the inverse of the Fisher information, which in turn quantifies the information carried by an observable about an unknown parameter (upon which the observable depends)[5]. In optical metrology with coherent light, Fisher information is not merely a statistical construct; it can be understood as a quantity carried by electromagnetic waves as they propagate - it can resonate, diffract, and interfere[6,7]. Here, we demonstrate that the surroundings of a measurement target can be engineered to improve the flow of Fisher information to a detector, thereby enhancing the measurement precision.

As a test case, we consider optical scattering-based measurements of the lateral position/displacement of a subwavelength width nanowire suspended within a gap in an opaque screen (Fig. 1)[8,9]. The screen is illuminated with coherent light, and transmission scattering patterns at a reference plane a few wavelengths from the screen are recorded via a high-numerical-aperture lens. (The scattered field in the reference plane is formed by the superposition of free-space propagating waves and is projected onto an image sensor without loss of resolution.) A neural-network estimator is trained on a large set of scattering patterns corresponding to known nanowire positions, enabling subsequent retrieval of unknown positions from previously unseen single-shot scattering patterns.

In the Fisher information (FI) framework, a change in the nanowire's lateral position creates a pair of coherent information sources – one at each side of the nanowire, where the material and/or field changes when the nanowire moves[6]. The flow of energy in an electromagnetic field is described by the Poynting vector $\boldsymbol{S_P} = \frac{1}{2}Re(\boldsymbol{E}^* \times \boldsymbol{H})$, where $\boldsymbol{E}$ and $\boldsymbol{H}$ are the electric and magnetic fields. The flow of information is analogously given by: $\boldsymbol{S_{FI}} = \frac{2}{\hbar\omega}Re(\partial_\theta \boldsymbol{E}^* \times \partial_\theta \boldsymbol{H})$, where $\partial_\theta \boldsymbol{E}$ and $\partial_\theta \boldsymbol{H}$ are 'sensitivity fields', which describe the response of the electric and magnetic fields to a change in the parameter $\theta$ (in our case, the lateral position of the nanowire). In systems with constant permeability $\mu_0$, FI flow is fully described by the electric sensitivity fields alone, which satisfy the wave equation $\nabla \times \nabla \times (\partial_\theta \boldsymbol{E}) + \mu_0 \epsilon (\partial_t^2 [\partial_\theta \boldsymbol{E}]) = \boldsymbol{Q}$, where $\boldsymbol{Q} = -\mu_0 (\partial_t \boldsymbol{E})(\partial_\theta \epsilon)$ is a source term that is non-zero only where $\epsilon$ and/or $\boldsymbol{E}$ vary with $\theta$.

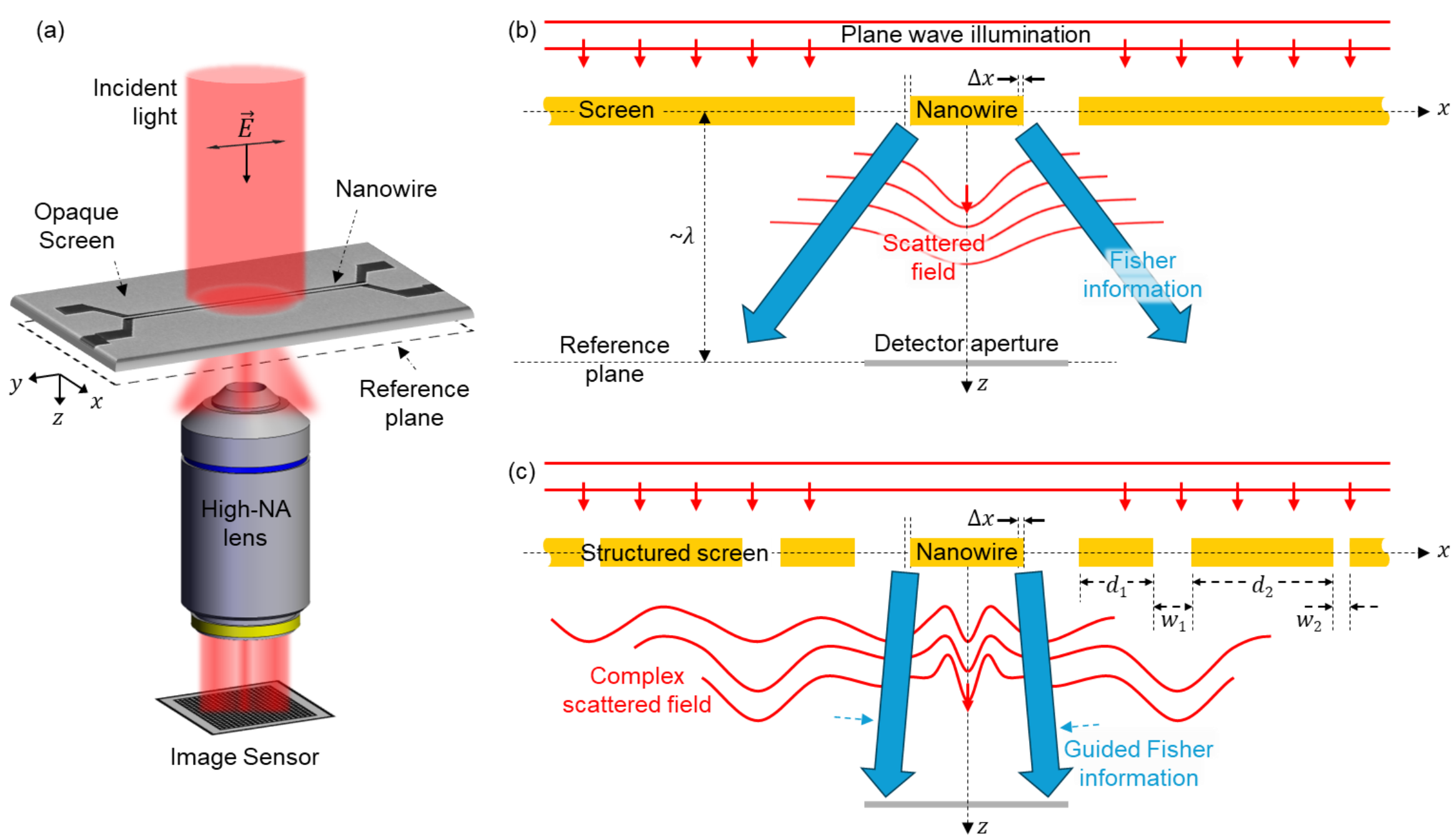

**Figure 1: Controlling the flow of Fisher information in optical scattering metrology.** (a) Schematic of experimental apparatus for optical scattering-based measurement of in-plane [$x$-direction] nanowire displacement. (b, c) Comparison of Fisher information flows for lateral $\Delta x$ displacement of a nanowire [illustrated in $xz$ cross-section]: (b) in a gap in an otherwise unstructured screen; (c) in and a structured screen where it is flanked [symmetrically in $\pm x$ directions] by pairs of parallel slits.

$\partial_\theta \boldsymbol{E}$ is not constrained to coincide with $\boldsymbol{E}$, and thus FI flow does *not* generally coincide with optical energy flow. In the case of a nanowire undergoing lateral displacement, the two FI sources are of equal magnitude but opposite phase (because $\partial_\theta \epsilon$ is positive at one side and negative at the other), leading to destructive FI interference along the $x = 0$ axis of the system[7]. As such, a detector centred on the optical axis – where transmitted energy is typically maximised – will capture only a small fraction of the available FI (Fig. 1b), thereby limiting achievable measurement precision. In this work, we show that an information-based inverse design approach can be applied to the surroundings of the measurement target (the structure of the screen around the nanowire – Fig. 1c) to maximise the fraction of generated FI reaching the detector, and so to enhance measurement precision.

Photodetectors and cameras do not detect FI – they record light intensity $I = |\boldsymbol{E}|^2$; and changes in light intensity do not correspond to the magnitude of the sensitivity field ($\partial_\theta I \neq |\partial_\theta \boldsymbol{E}|^2$). The Fisher information obtained from intensity measurements is given by $FI = (\partial_\theta I)^2 / I \sim u^{\mathrm{FI}} \cos^2 \Delta\phi$, where $u^{\mathrm{FI}} = (\hbar\omega)^{-1} \epsilon |\partial_\theta \boldsymbol{E}|^2$ is the Fisher information density and $\Delta\phi$ is the phase difference between the sensitivity field and the scattered field. Optimising the environment of a measurement target thus amounts to engineering the amplitude and phase of a scattered 'reference field' to maximise the proportion of FI accessible from intensity-only scattered field measurements.

In experiment, this approach to optimally structuring the surroundings of the nanowire – requiring the addition of just three pairs of parallel slits in the surrounding screen – yields a 6.5× improvement in positional measurement precision relative to the unstructured case: reducing the mean standard deviation ($\sigma = \sqrt{Var(x)}$) from 397 to 61 pm.

## Control of Fisher information flow with nanostructures

In numerical simulations using the angular spectrum method, we first consider a 200 nm wide nanowire centred within a 400 nm gap in an otherwise opaque screen, under normally incident plane wave illumination at a wavelength $\lambda$ = 633 nm (see Methods). From the scattered field (in transmission) we evaluate the Fisher information associated with small lateral displacements of the nanowire ($|\Delta x|$ = 1 nm), integrated over a central 2 μm window in the reference plane, representing the back-projected finite aperture of a detector. A Bayesian optimisation procedure[10] is then employed to maximise detected FI by varying the widths $w_i$ and nearest-edge separations $d_i$ of $N$ additional pairs of slits in the screen, positioned symmetrically on either side of the nanowire, as illustrated in Fig. 1b. The optimisation is performed under constraints that $w_i \geq$ 50 nm and $d_i \leq$ 2 μm, ensuring that the parameter space is reasonably sized, and features remain practical from a nanofabrication perspective.

The presence of a single optimised pair of additional slits in the screen enhances detected FI by 39×; with three or more pairs of slits this enhancement factor reaches almost 50×. (Optimisation learning curves for screens containing up to four additional slit pairs are shown in Supplementary Fig. S1.) The mechanisms by which an optimally structured screen maximises detected Fisher information can be understood from maps (Fig. 2) and reference plane profiles (Fig. 3) of the scattered field and FI. In the absence of any additional slits in the screen, most (90%) of the energy transmitted by the gap around the nanowire is detected because it propagates along the $x = 0$ axis of the system, directly towards the detector (Fig. 2a). However, very little of the available Fisher information is detected because, as discussed above, information associated with lateral $\Delta x$ displacements of the nanowire interferes destructively along the $x = 0$ axis; it is scattered at high angles, beyond the detector's ±~38° acceptance angle in the present case (Fig. 2c).

As additional slits are introduced, the intensity and phase structure of the scattered field becomes increasingly complex, and the distribution of Fisher information about the $x = 0$ axis becomes narrower. In the optimal case with four additional pairs of slits ($N$ = 4, Fig. 2d-f), detected power is 6.3× higher and detected FI is ~50× higher than in the unstructured case. This corresponds to an enhancement of Fisher information per unit detected power (per photon) by a factor of 7.8. It is important to note here that while total power transmitted through the screen increases with the number of additional slits, the amount of Fisher information generated does not change: the information sources at the edges of the nanowire are independent of the surrounding screen structure.

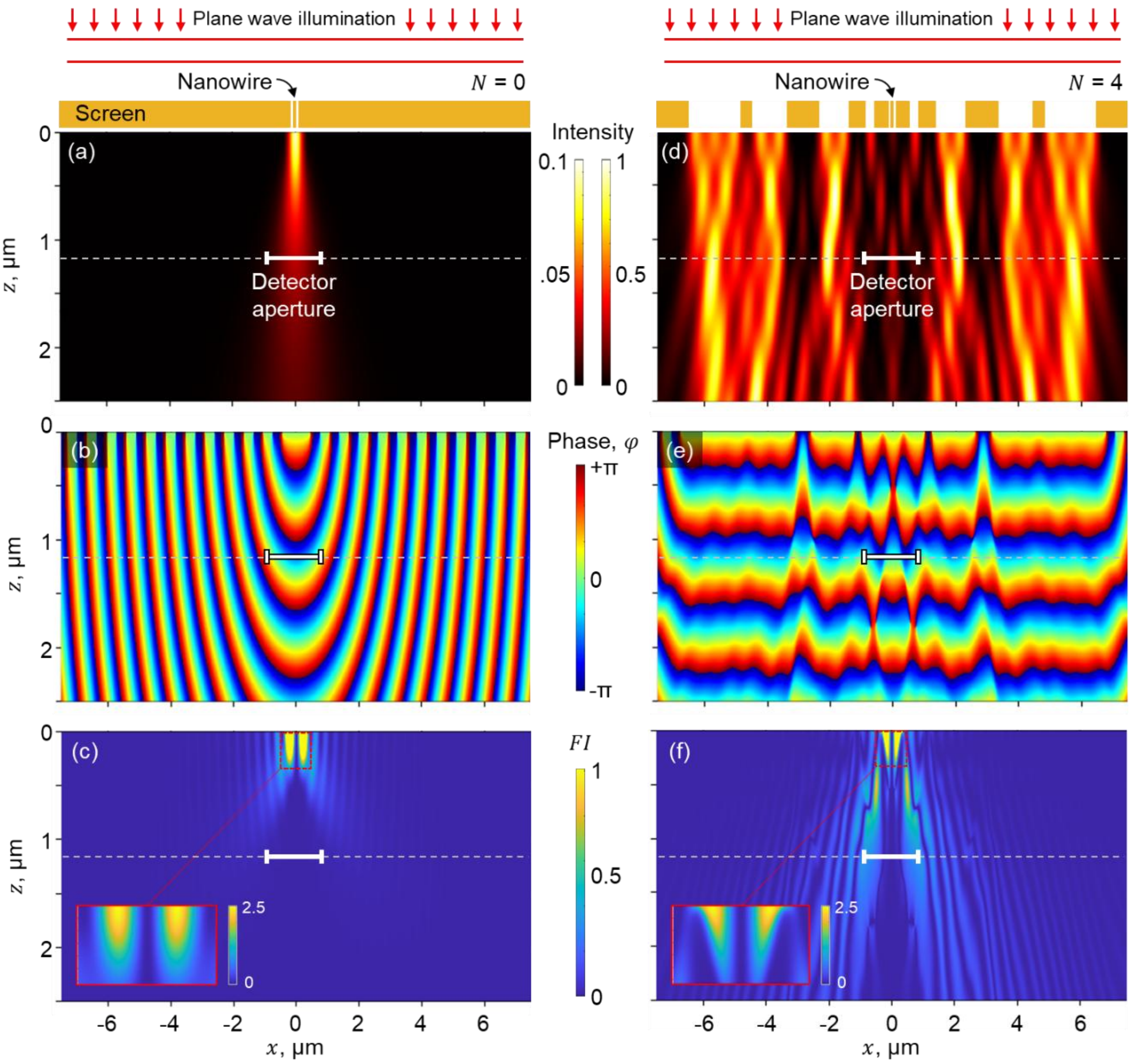

**Figure 2: Maximising accessible Fisher information by optimising the target environment.** Maps of transmitted field intensity $I$ and phase $\varphi$, and corresponding Fisher information $FI \propto (\partial_\theta I)^2/I$ for a nanowire in: (a-c) an unstructured screen; (d-f) a structured screen with 4 additional pairs of slits, the positions and widths of which have been optimised with respect to detected FI. [Corresponding maps for structured screens with 1-3 additional pairs of slits are presented in Supplementary Fig. S2.]

Interference between light scattered from the nanowire (through the gaps on either side, which change size as the nanowire moves) and light transmitted through the additional (fixed) slits structures the reference-plane field at the subwavelength scale (Fig. 3). Broad, single-peak (or trough) intensity and phase profiles in the unstructured ($N = 0$) case are replaced by oscillatory profiles with multiple intensity and phase-gradient zero-crossings across the detector aperture. In the optimal case with four pairs of additional slits, the field profile contains a central sub-diffraction limit hotspot (with a half-maximum width of $\lambda/2.2$ – Fig. 3a), flanked by regions where the local phase gradient exceeds the free-space bandlimit – characteristic features of superoscillatory fields (Fig. 3c). Such topological features generally arise from subtle interference effects[11,12], and are consequently highly sensitive to perturbations at their source – i.e., in this case, to changes in the nanowire position relative to the static, structured screen.

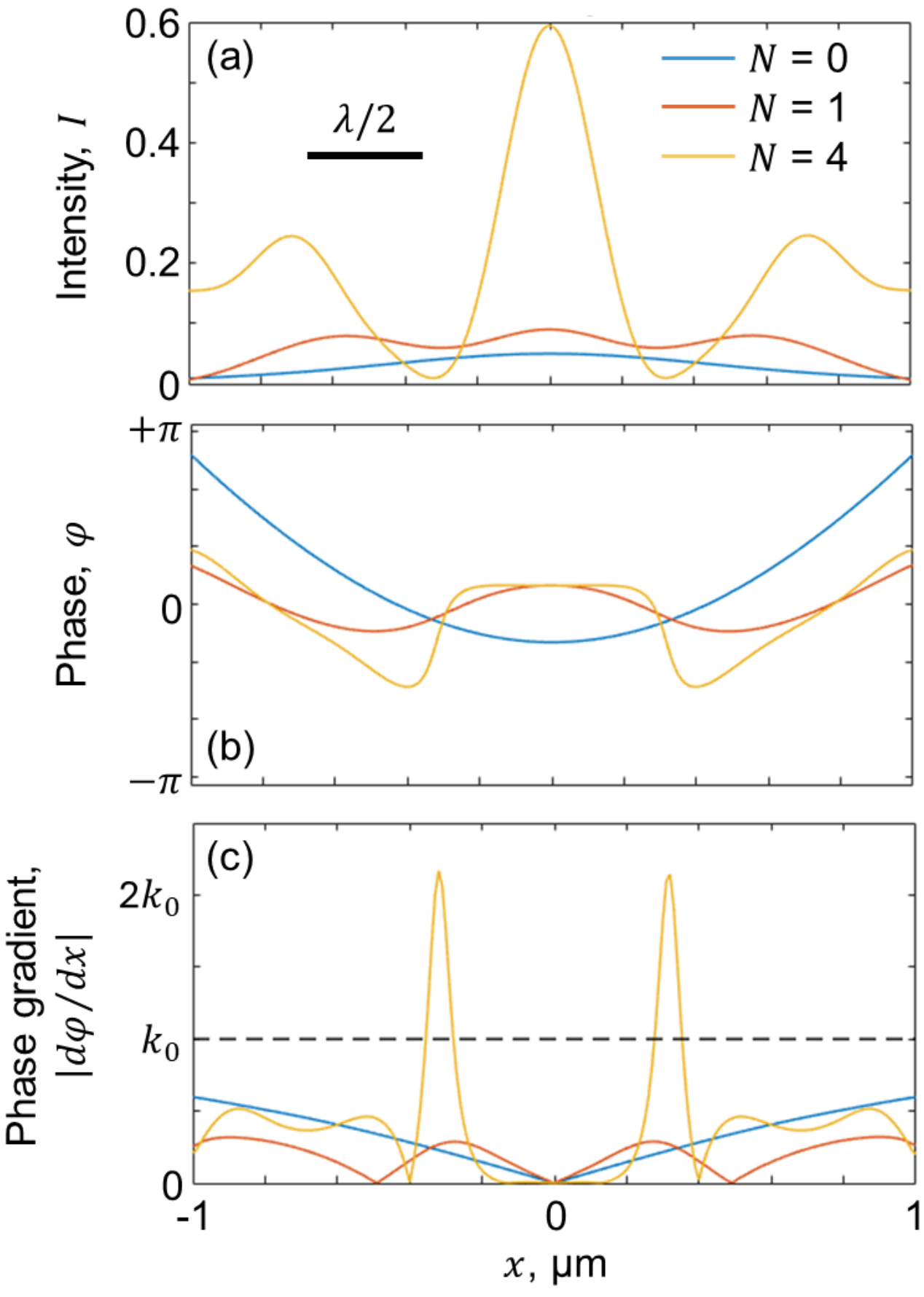

**Figure 3: Subwavelength structure of the scattered field**. Profiles of: (a) intensity; (b) phase; (c) and phase gradient of the scattered field in the reference plane, over the 2 μm detector aperture indicated in Fig. 2, for an unstructured screen and screens optimally structured with 1 pair and 4 pairs of additional slits. [$k_0$ is the free-space wavevector.]

**Information-driven optimisation of localisation measurements**

To generate optimal structured screen designs for experimental nanowire samples, we apply the Fisher information-based optimisation procedure (as described above) to slit widths and separations in 3D finite-difference time-domain (FDTD) simulations of a free-standing, gold-coated silicon nitride screen (as used in prior experimental studies[8,9] – 65 nm Au on 50 nm $Si_3N_4$ – see Methods). Three pairs of additional slits yield a maximum FI enhancement of ~28× (see Supplementary Table ST1), whereby measuremnt precision is expected to improveby √28 ~5.3×. These numbers differ from those obtained via the angular spectrum method above (where FI enhancement reaches ~50× with three pairs of slits) because the FDTD model accounts for absorption losses (in gold) and retardation effects due to finite thickness of the materials. We also note that, in the presence of surface plasmon polaritons excited on gold surfaces, and scattered light propagating through (and/or reflected within) the silicon nitride layer, the amount of FI generated is not strictly invariant to changes in the nanowire environment: the presence of additional slits in the screen may enhance or reduce the magnitude of the electric field at the edges of the nanowire, and thereby the magnitude of the information sources.

Experimental measurements of $\Delta x$ lateral displacement were performed on nanowires flanked by up to three pairs of additional slits (Fig. 4a, b, and Supplementary Fig. 4a-c), at a wavelength $\lambda$ = 640 nm, using a neural network estimator to retrieve nanowire positions from single-shot transmission scattering patterns[8,9] (see Methods). The substantial improvement in measurement performance arising from the presence of additional slits is clearly seen in Fig. 4c, which plots optically measured versus actual nanowire displacement for the unstructured-screen ($N$ = 0) reference case and that of a screen optimally structured with $N$ = 3 pairs of additional slits. The spread of points in each case is derived from 12 separate measurements at each actual displacement value (see Methods). Taking the mean measurement standard deviation $\sigma$ over the full range of displacement values as a figure of merit, three pairs of optimised slits deliver a 6.5× improvement over the unstructured case: $\sigma$ = 61 pm compared to 397 pm. This is slightly better than the factor of 5.3× expected from the numerically simulated level of $FI$ enhancement. This discrepancy may be attributed to a combination of: (i) imperfect matching between computational and experimental object-to-reference-plane distances – resulting in a detector acceptance angle that is slightly larger in experiment than assumed in simulations; (ii) the fact that numerical modelling does not account for fabrication imperfections (e.g. nanowire edge roughness over its illuminated length) or the real (defocused Gaussian as opposed to plane wave) incident wavefront – both of which may add information to scattering patterns that is beneficial to the neural network's training and measurement retrieval process.

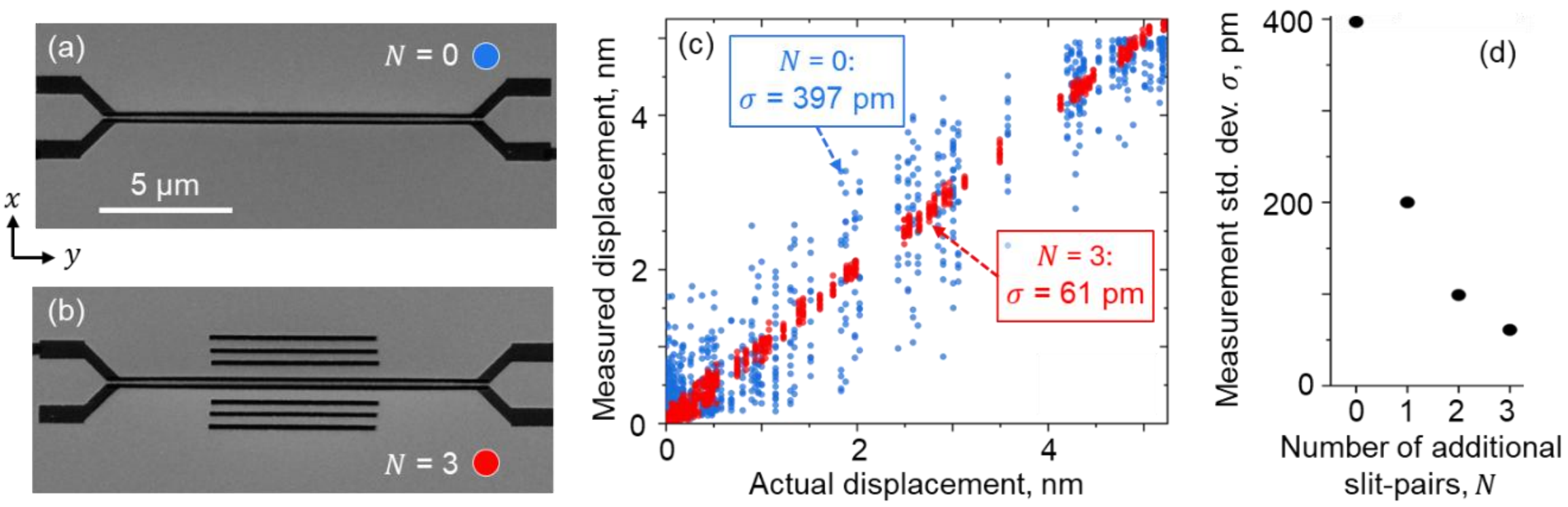

**Figure 4**: **Improving the precision of localisation measurements by optimising the target environment.** (a, b) Scanning electron microscope images of nanowire samples: (a) surrounded by an unstructured screen; (b) in a screen optimally structured with three pairs of additional slits. (c) Optically measured versus actual values of nanowire displacement for the samples shown in (a, b). [Corresponding data for samples with one and two pairs of additional slits in the screen are presented in Supplementary Fig. S4.] (d) Mean standard deviation of measurements over the 0 - 5.25 nm displacement range, as a function of the number of additional slit-pairs [with optimised positions and widths].

**Conclusion**

In summary, we demonstrate that in optical scattering metrology, structuring the surroundings of a measurement target can yield a multi-fold improvement in precision. In single-shot nanowire localisation experiments at a wavelength of 640 nm, this approach reduces the standard deviation of measurements from ~400 pm to ~60 pm through the introduction of additional scattering elements designed to maximise the proportion of Fisher information reaching a finite aperture detector. Interference between light scattered by the target and by the additional elements yields an intensity distribution at the detector that is maximally sensitive to the measurand. This work exemplifies that the Fisher information flow can be controlled in much the same way as energy flow by structuring the electromagnetic environment.

**Methods**

Simulations using the angular spectrum method[13] are performed in 2D, assuming a perfectly reflecting, non-absorbing screen (and nanowire in cross-section) of vanishing thickness in the $z$ (incident light propagation) direction. The detector aperture, at a distance $z = 2\lambda$ beyond the screen, is taken to have a width $L_d$ = 2 μm and a 5 nm effective pixel size (in keeping with experiments – see below). We assume detector performance to be shot noise-limited, which implies that the measured intensity at each pixel is governed by a Poissonian noise distribution.

Finite difference time domain numerical simulations (in Ansys Lumerical FDTD) use material parameters for gold and silicon nitride from Johnson and Christy, and Phillipp, respectively[14,15]. They assume normally-incident plane wave illumination at a wavelength $\lambda$ = 640 nm. $FI$ is evaluated from scattered fields over a 3 μm × 3 μm area in the $xy$ plane at a distance $z$ = 0.5 μm from the back of the screen, with an effective pixel size of 5 nm. The modelling domain has dimensions of 4 μm, with periodic boundary conditions, in the $y$ direction parallel to the length of the nanowire (i.e. effectively assuming an infinitely long nanowire); 10 μm in the $x$ direction, centred on the nanowire with perfectly-matched layer (PML) boundary conditions; and 1 μm in the $z$ direction, centred 0.3 μm from the back of the screen, with PML boundary conditions.

Experimental samples – 17 μm long nanowires flanked by up to 3 pairs of additional slits – were fabricated by focused ion beam milling on a 50 nm silicon nitride membrane coated by evaporation with 65 nm of gold.

Optical measurements of nanowire displacement were performed by illuminating samples, at the mid-point of the nanowire length, with a defocused Gaussian beam (spot diameter >50 μm, $\lambda$ = 640 nm) polarized perpendicular to the nanowire length, via a 20×, NA 0.4 objective. Transmission scattering patterns from a reference plane 0.5 μm from the sample were recorded on a 12-bit sensor with a $13\lambda \times 13\lambda$ (250 × 250 pixel) field of view, via a 100×, NA 0.9 objective and 3× tube lens (not shown in Fig. 1a). For each sample ($N$ = 0-3), twelve consecutive sets of scattering patterns for 400 different (electrostatically controlled) nanowire positions over a range of $\Delta x$ displacements up to 5.25 nm from the zero-bias position – a total of 4800 scattering patterns (captured over a period of 48 ms), with each nanowire position revisited 12 times (at intervals of ~4 ms).

The twelve scattering patterns corresponding to each of 320 randomly-selected 'known' nanowire positions (80% of the 400 available) were used for neural network training and validation; the remaining 20% were reserved for testing, i.e. as unseen patterns for nominally unknown displacements.

The *in-situ* training regime, i.e. whereby the estimator is trained and tested on the same (variable) object, rather than a large set of discrete objects, maximises congruence between training and testing data[8]. Importantly, the iteration of known nanowire positions within the training set enables the estimator to distinguish between sub-nanometric displacements of the nanowire relative to the edges of the gap in the screen, and (much larger) movements of the whole sample due to instrumental fluctuations[9]. Full details of neural network architecture and nanowire displacement calibration (dataset labelling) procedure are given in Refs. [8, 9].

**Acknowledgements**: This work was supported by the UK Engineering and Physical Sciences Research Council (grants EP/T02643X/1 and EP/Z53285X/1). The authors would like to acknowledge discussions with Huanli Zhou and Eric Plum (University of Southampton, UK), Maximilian Weimar, Luca Neubacher, Jakob Hüpfl, and Stefan Rotter (TU Wien, Austria), which informed the development of the manuscript.

## Improving Optical Metrology by Engineering the Target Environment - Supplementary Information

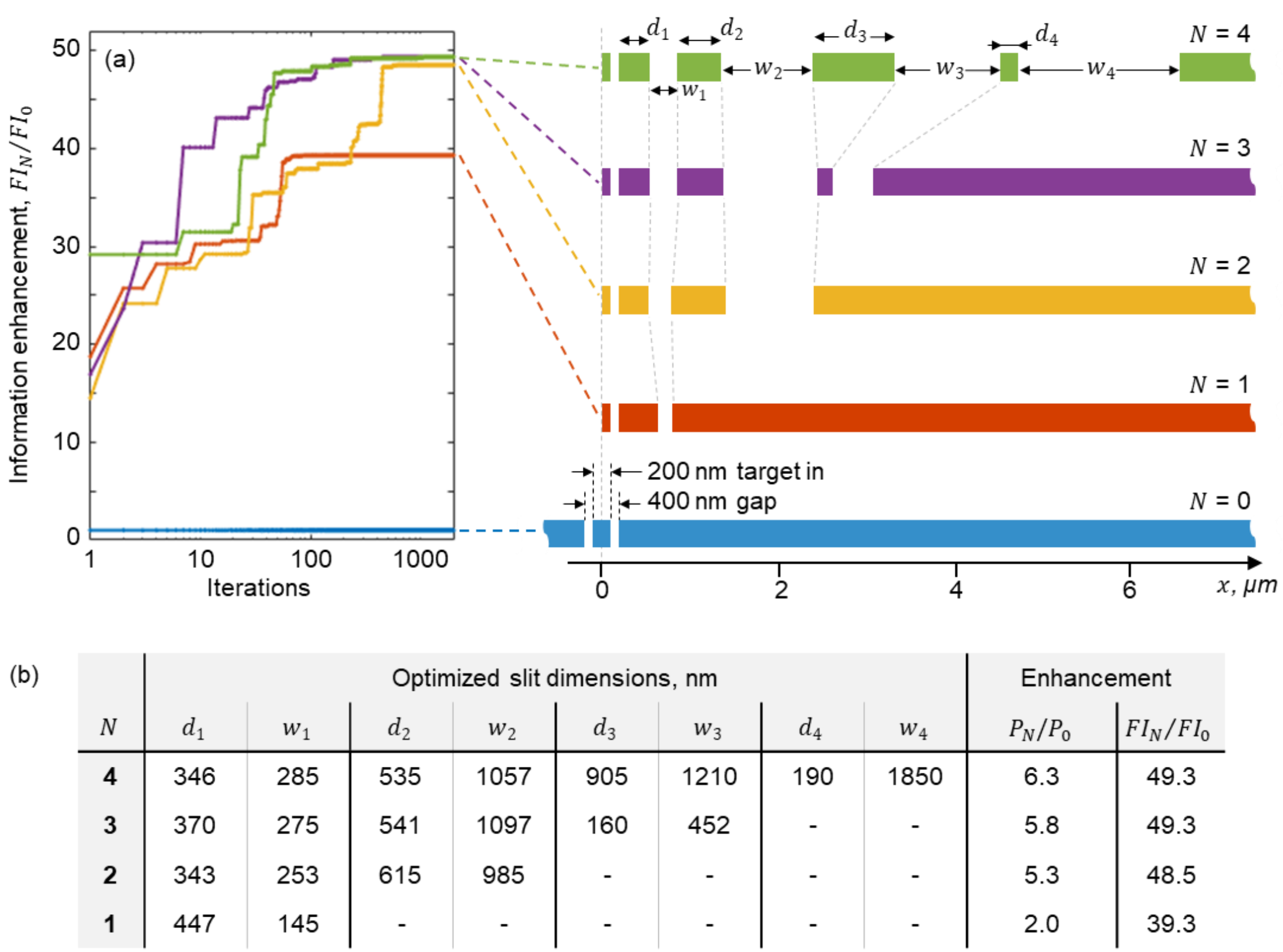

| $N$ | Optimized slit dimensions, nm | | | | | | | | Enhancement | |
|---|---|---|---|---|---|---|---|---|---|---|
| | $d_1$ | $w_1$ | $d_2$ | $w_2$ | $d_3$ | $w_3$ | $d_4$ | $w_4$ | $P_N/P_0$ | $FI_N/FI_0$ |
| **4** | 346 | 285 | 535 | 1057 | 905 | 1210 | 190 | 1850 | 6.3 | 49.3 |
| **3** | 370 | 275 | 541 | 1097 | 160 | 452 | - | - | 5.8 | 49.3 |
| **2** | 343 | 253 | 615 | 985 | - | - | - | - | 5.3 | 48.5 |
| **1** | 447 | 145 | - | - | - | - | - | - | 2.0 | 39.3 |

**Figure S1: Optimising target environment for lateral displacement measurements.** (a) Learning curves for the iterative process of maximising detected Fisher information [over a finite detector aperture as illustrated in Fig. 1c] associated with $\Delta x = \pm 1$ nm displacements of a 200 nm wide nanowire centred within a 400 nm gap in an otherwise opaque screen, under normally incident plane wave illumination at a wavelength $\lambda = 633$ nm, through the introduction of up to $N = 4$ pairs of additional slits [of variable width $w_i$ and separation $d_i$, symmetric about $x = 0$] in the screen. Curves show detected FI enhancement relative to the unstructured [$N = 0$] reference case, as a function of iteration number. Optimal screen profiles (for the $x > 0$ half-space) are illustrated to the right. (b) Optimal slit dimensions, and corresponding detected power ($P$) and FI enhancement factors.

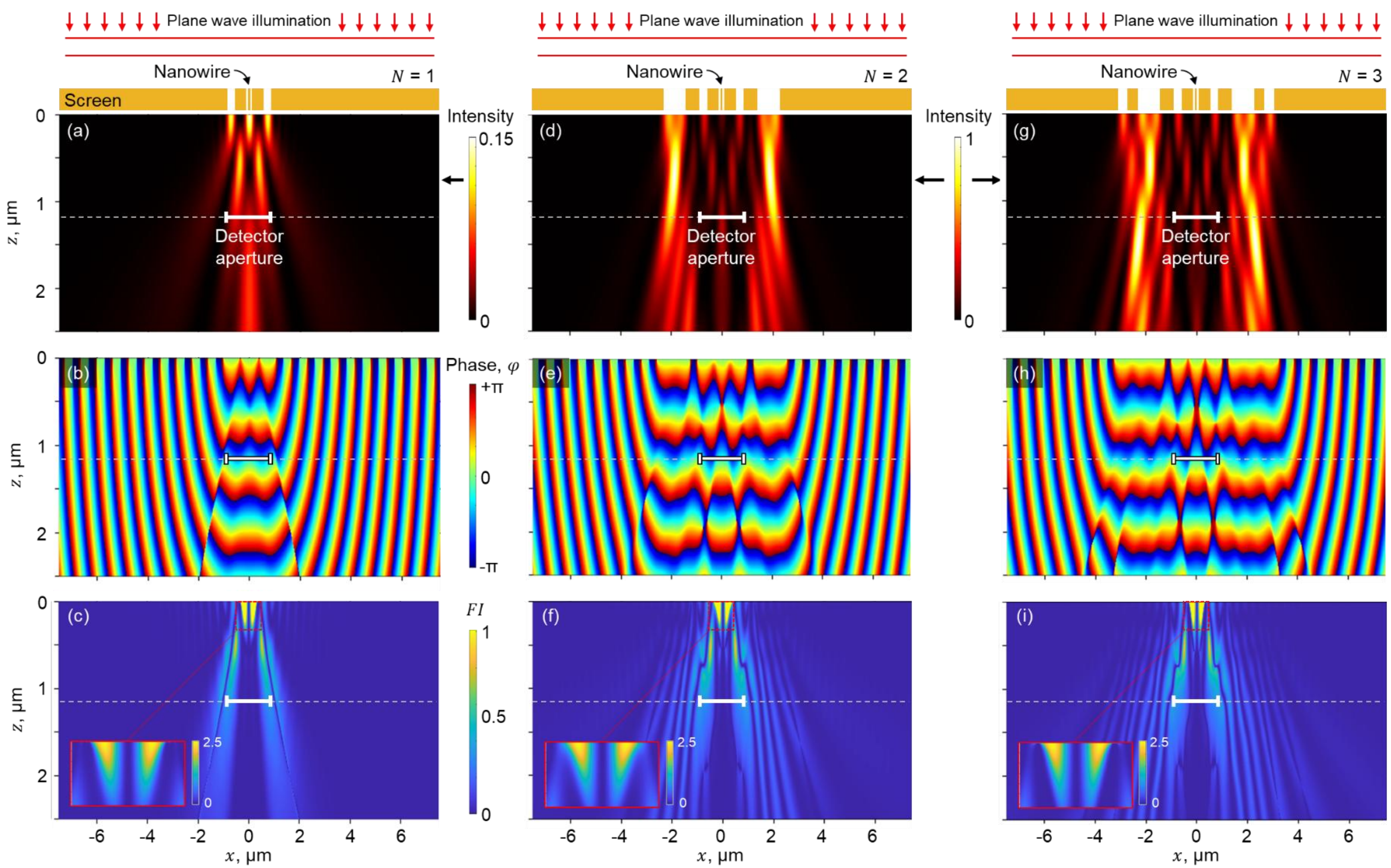

**Figure S2: Scattered field and Fisher information.** Maps of transmitted field intensity $I$ and phase $\varphi$, and corresponding Fisher information $FI \propto (\partial_\theta I)^2/I$ for a nanowire in structured screens with: (a-c) $N = 1$; (d-f) $N = 2$; and (g-i) $N = 3$ additional pairs of slits, , the positions and widths of which have been optimised with respect to detected FI. [Note that the intensity colour scale maximum value in panel (a) is only 15% of the value in panels (d) and (g).]

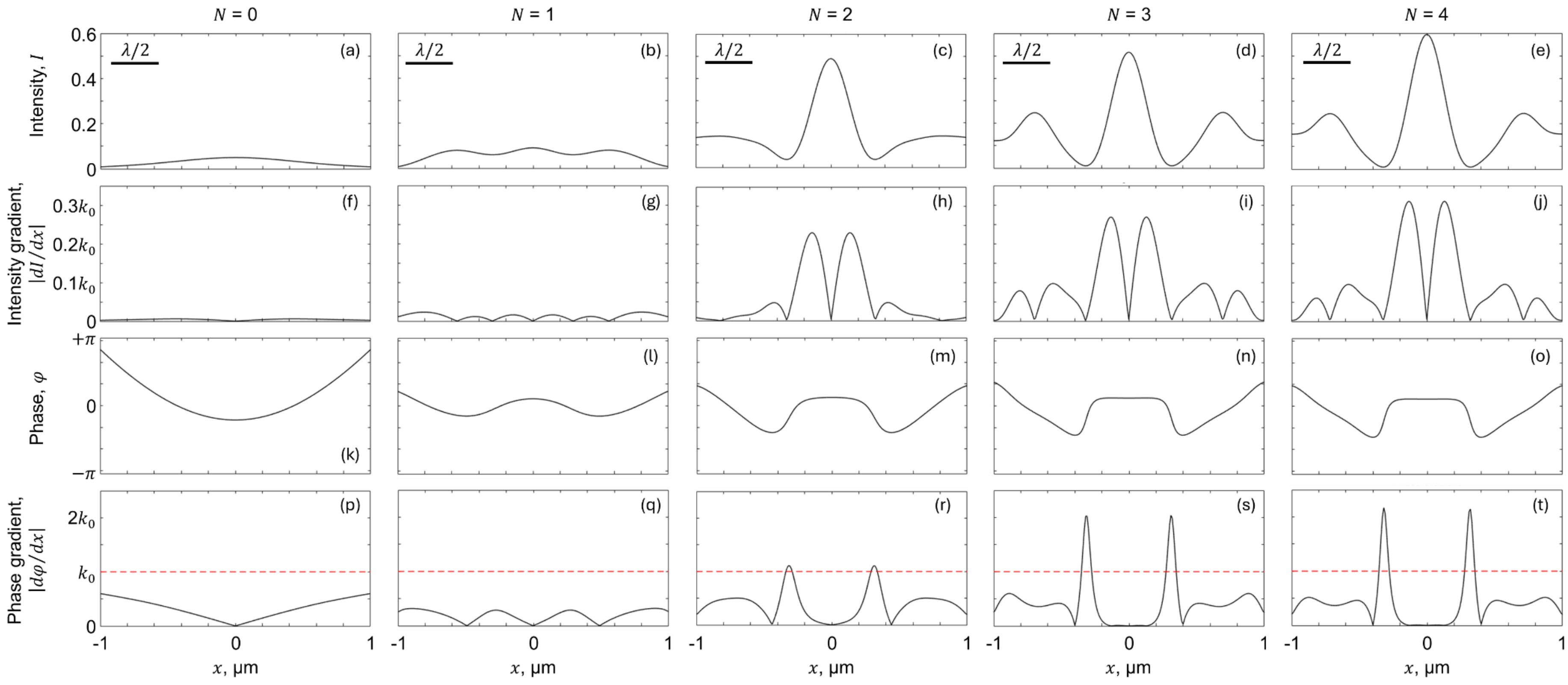

**Figure S3: Reference plane field profiles.** (a-e) Intensity, (f-j) intensity gradient, (k-o) phase, and (p-t) phase gradient of in the reference plane, over the 2 μm detector aperture indicated in Fig. 2, for an unstructured screen [$N$ = 0] and screens optimally structured with $N$ = 1-4 additional pairs of slits [as labelled, from left to right].

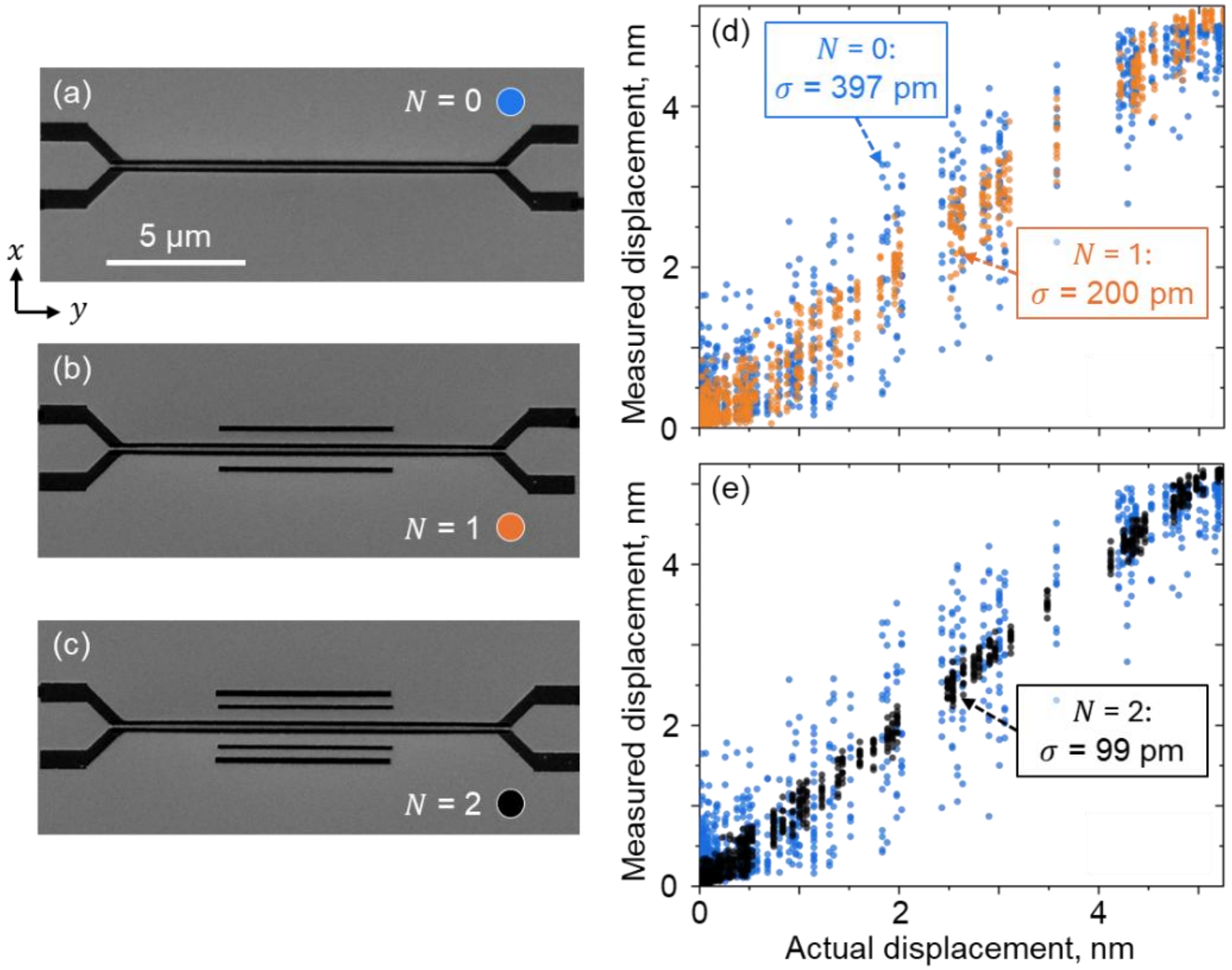

**Figure S4: Improving the precision of localisation measurements by optimising the target environment.** (a-c) Scanning electron microscope images of nanowire samples: (a) surrounded by an unstructured screen; (b, c) in screens optimally structured with one and two pairs of additional slits. (d, e) Optically measured versus actual values of nanowire displacement for the samples shown in (a-c). The spread of points in each case is derived from 12 separate measurements at each actual displacement value.

**Table ST1: Optimal slit dimensions (nm) for experimental (gold-coated silicon nitride) samples, and corresponding Fisher information enhancement factors, from 3D finite element modelling.**

| $N$ | $d_1$ | $w_1$ | $d_3$ | $w_2$ | $d_3$ | $w_3$ | Enhancement $FI_N/FI_0$ |
|---|---|---|---|---|---|---|---|
| **3** | 580 | 200 | 360 | 180 | 410 | 200 | 28 |
| **2** | 580 | 160 | 400 | 240 | - | - | 27 |
| **1** | 570 | 210 | - | - | - | - | 9.1 |